\documentclass[%
 aip,
 amsmath,amssymb,
 reprint,%
]{revtex4-1}
\usepackage{float}
\usepackage{graphicx}
\usepackage{dcolumn}
\usepackage{bm}
\usepackage{subcaption}
\usepackage{xcolor}
\usepackage[utf8]{inputenc}
\usepackage[T1]{fontenc}
\usepackage{mathptmx}
\usepackage{etoolbox}
\usepackage{graphicx}

\makeatletter
\def\@email#1#2{%
 \endgroup
 \patchcmd{\titleblock@produce}
  {\frontmatter@RRAPformat}
  {\frontmatter@RRAPformat{\produce@RRAP{*#1\href{mailto:#2}{#2}}}\frontmatter@RRAPformat}
  {}{}
}%
\makeatother
\begin{document}

\title[]{High-Field Electron Transport in AlGaN alloys: A Full-Band Monte Carlo Study Based on Ab Initio Supercell Calculations}
\author{Animesh Datta}
 \author{Matinehsadat Hosseinigheidari}
\author{Uttam Singisetti}%
 \email{uttamsin@buffalo.edu}
\affiliation{ 
Department of Electrical and Computer Engineering, University at Buffalo (SUNY), Buffalo, New York 14228, USA
}%
\date{\today}

\begin{abstract}
 AlGaN alloys are promising wide and ultra-wide-bandgap semiconductors for next-generation power and RF electronics applications. To realize the full potential of AlGaN based devices, it is important to understand the electron transport to accurately predict device performance and identify material limits under various operating conditions. In this work, the high-field electron transport properties of Al$_x$Ga$_{1-x}$N are investigated using a supercell-based full-band Monte Carlo method. The supercell approach is employed to explicitly capture the true disorder of the alloy system, enabling a more realistic description of carrier transport. The velocity–field characteristics are calculated across a range of Al compositions to evaluate key transport metrics, including peak velocity, saturation velocity, and critical electric field. The role of different scattering mechanisms is studied in detail to understand the high-field transport mechanism in the AlGaN alloy system. In addition to steady-state transport, transient electron dynamics are examined for various Al fractions to study velocity-overshoot behavior, which is especially important for improving the performance of scaled RF devices. Finally, the temperature dependence of the velocity–field characteristics in ultra-wide-bandgap Al$_{0.75}$Ga$_{0.25}$N is investigated to assess its transport performance under high temperature conditions. These results provide a detailed understanding of high-field transport in AlGaN alloys and offer guidance for the design of AlGaN-based RF and power electronic devices.
\end{abstract}

\maketitle

\section{\label{sec:level1}Introduction and Overview}
In the last decade, wide- and ultra-wide-bandgap semiconductors have gained significant attention due to the escalating demand for higher power density, improved energy efficiency, and reliable operation at elevated voltages and temperatures, particularly in emerging applications such as electric vehicles, renewable energy systems, and millimeter-wave communication technologies\cite{aminbeidokhti2011novel,murugapandiyan2020gan,ueda2014gan}. Among these materials, group III nitrides possess a highly attractive combination of wide and tunable band gaps\cite{vurgaftman2001band}, high breakdown fields, strong polarization effects\cite{bernardini1997spontaneous}, and excellent suitability for both power and high-frequency device applications\cite{hoo2021emerging,kaplar2017ultra}. In particular, alloy systems like AlGaN with their tunable properties provide an expanded design space for various power and RF applications. 

In the early years, AlGaN was mainly used as a barrier layer for AlGaN/GaN HEMT devices and achieve a high density two dimensional electron gas (2DEG) for radio-frequency (RF) electronics. \cite{mishra2002algan,mishra2008gan,klein2024rich,baca2016aln,anderson2009aln,douglas2019enhancement,xue20200,moon2021power}. However in recent years, there has been substantial interest in using AlGaN as the active channel layer for UWBG electronic devices. The motivation behind using AlGaN as the channel layer is to achieve useful power gain at high power density\cite{xue2019al0,xue20200,ye2022electron,singhal2022toward,kim2023rf}. In RF electronics,  device performance is often quantified using Johnson’s Figure of Merit (JFoM) defined as JFoM = $\frac{E_c v_s}{2\pi}$ where $v_s$ is the saturation velocity. Earlier experimental and theoretical studies have characterized the velocity–field behavior in bulk GaN and AlN\cite{fang2019electron,yamakawa2009rigid,albrecht1998monte,o2006steady,foutz1999transient,dyson2015hot,bertazzi2009theory,barker2002high,collazo2003electron}; however, for bulk AlGaN alloys, experimental reports remain unavailable. Thus it is important to investigate and understand the velocity-saturation effects in AlGaN to optimize and realize the full potential of AlGaN based devices. 
\\

Prior theoretical investigations of high-field transport in AlGaN are based on the nonlocal empirical pseudopotential method (NL-EPM) and the virtual crystal approximation (VCA) method to study the velocity–field characteristics \cite{farahmand2001monte,bellotti2007alloy,coltrin2017transport}. Recently, supercells (SC) approaches have gained prominence for explicitly capturing the true disorder of the alloy systems by reducing the defect-defect interactions \cite{popescu2010effective,popescu2012extracting}. Using SC, researchers have explored the effective band structure and phonon dispersion of semiconductor alloys (\textit{e.g. }AlGaO, AlGaN, ScAlN) to gain critical insights into the thermal conductivity, sound velocity and alloy scattering potential \cite{balestra2022electron,kyrtsos2019first,datta2024effective,sharma2023effective,pant2020high}. In previous works, we have also explored the low-field transport properties in alloy systems like AlGaO, AlGaN using SC \cite{datta2024effective,sharma2025effective}. However there is no report of a first principles based high-field transport study in AlGaN alloys using SC.

Thus, in this work, we employ SC electronic structures combined with full-band \textit{ab-initio} Monte Carlo simulations to investigate the high-field transport mechanisms in AlGaN alloys. In a  previous report \cite{datta2024effective}, we performed the Density Functional Theory (DFT) calculations and the Density Functional Perturbation Theory (DFPT) calculations on a 24-atom SC AlGaN and obtained the required phonon and electronic band structure properties. Using those parameters in this work, we first calculate the full band deformation potential and long range scattering rates using the Fermi Golden Rule. Then using the Full Band Monte Carlo (FBMC) method outlined in Section II, we obtain the transient dynamics and velocity field characteristics in AlGaN for various Al fractions. From the velocity-field profiles and the energy distribution of electrons, the underlying mechanism of high-field transport in the AlGaN alloys is studied in detail to understand the effects of various scattering mechanisms. The transient dynamics of the AlGaN systems is compared for the three Al fractions ( 0.25, 0.5 and 0.75) and investigated in detail to enable the full potential of the material for RF applications. Finally, we also studied the effect of temperature on the velocity-field characteristics for understanding high temperature device operation of AlGaN based devices.

\section{Methodology and Computational Details}
The process of calculating the high-field transport in AlGaN alloys starts with the first principles calculation of the electronic and phonon bandstructures. Using the DFT and DFPT calculated parameters, the \textit{ab-initio} scattering rates are calculated using Fermi's Golden Rule. Then using the \textit{ab-initio} calculate band energies, phonon frequencies and the scattering rates, the Full Band Monte Carlo calculation is performed to simulate the electron transport in bulk AlGaN. In this section, first we discuss the computational details behind the DFT, DFPT, and the scattering rate calculation. This is followed by a brief discussion on the Full Band Monte Carlo method to simulate the high-field transport properties.
\subsection{First Principles Calculations}
\begin{figure}
\centering
\includegraphics[width=0.5\textwidth]{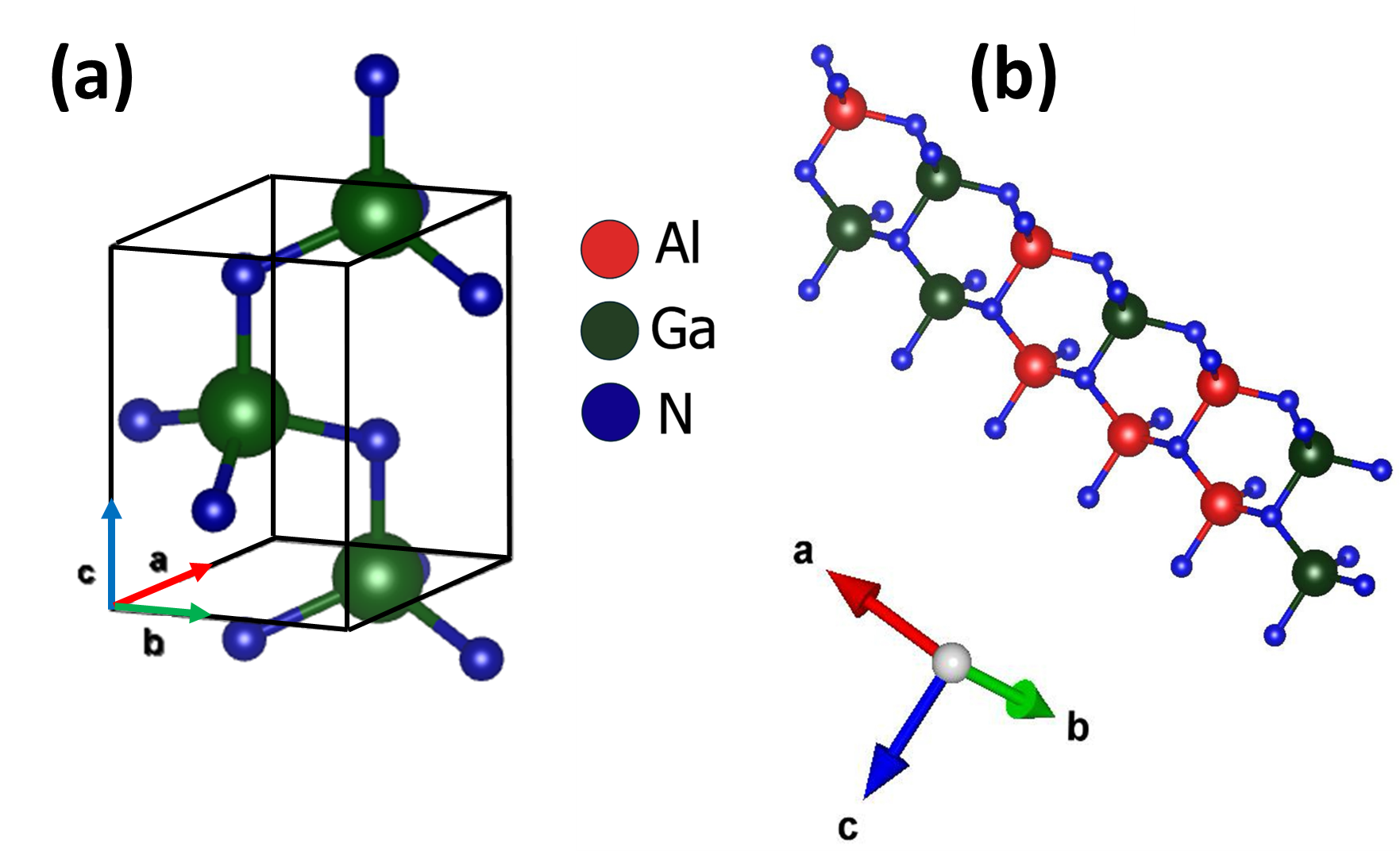}
\caption{(a) 4 atom GaN primtive cell (b) 24 atom Al$_{0.5}$Ga$_{0.5}$N supercell constructed using the SQS ATAT.}
\label{supercell}
\end{figure}

In this work, we have used a 24 atom supercell (SC) generated using the Alloy Theoretic Automated Toolkit (ATAT) \cite{van2009multicomponent,van2013efficient}. The SC was generated from the 4-atom PC as the fundamental repeating unit in the $6\times1\times1$ configuration as shown in Fig.\ref{supercell}.
As explained in our previous work \cite{datta2024effective}, the structure was first subjected to volume and atomic relaxations. Then using Quantum Espresso\cite{giannozzi2009quantum}, DFT and DFPT calculations were performed to obtain the electronic band-structure and the phonon dispersion. The electronic band-structure was then Wannier interpolated on a $40\times40\times40$ electronic wavevector (k) grid using the EPW package\cite{ponce2016epw, lee2023electron}. In our calculations we used 8 conduction bands  up to $\approx$ 3eV above the conduction band minima (CBM).
In the next step, the electron-phonon interaction (EPI) elements is calculated using the EPW package \cite{ponce2016epw, lee2023electron}. For the EPI calculation, the short range and the long range electron-phonon interaction elements are treated differently. The shortrange EPI elements are calculated on a $40\times40\times40$ k grid and a $20\times20\times20$ phonon waver-vector (q) grid using the EPW package through Wannier interpolation from  a coarse grid. The matrix elements are stored for each band transitions and each phonon mode. For the long range matrix elements, we use the formulation developed by by Verdi and
Giustino\cite{verdi2015frohlich}, which is a generalization of the Frohlich interaction as implemented in EPW. The matrix elements are calculated on a denser $40\times40\times40$ q grid to capture the stronger electron phonon interaction near the $\Gamma$ point. Using the Fermi Golden Rule and the calculated electron-phonon interaction elements, the polar optical phonon scattering, piezoelectric scattering and deformation potential scattering rates are calculated taking into consideration all the interband and intraband transitions. In addition, the ionized impurity scattering mechanism is modeled using the Brooks-Herring formulation\cite{chattopadhyay1981electron}. The scattering of alloy disorder is modeled using a statistically averaged disorder potential in the matrix element for Fermi’s golden rule \cite{bellotti2007alloy,pant2020high}.

\section{Full Band Monte Carlo}
\begin{figure}
\begin{center}
    \includegraphics[width=0.5\textwidth]{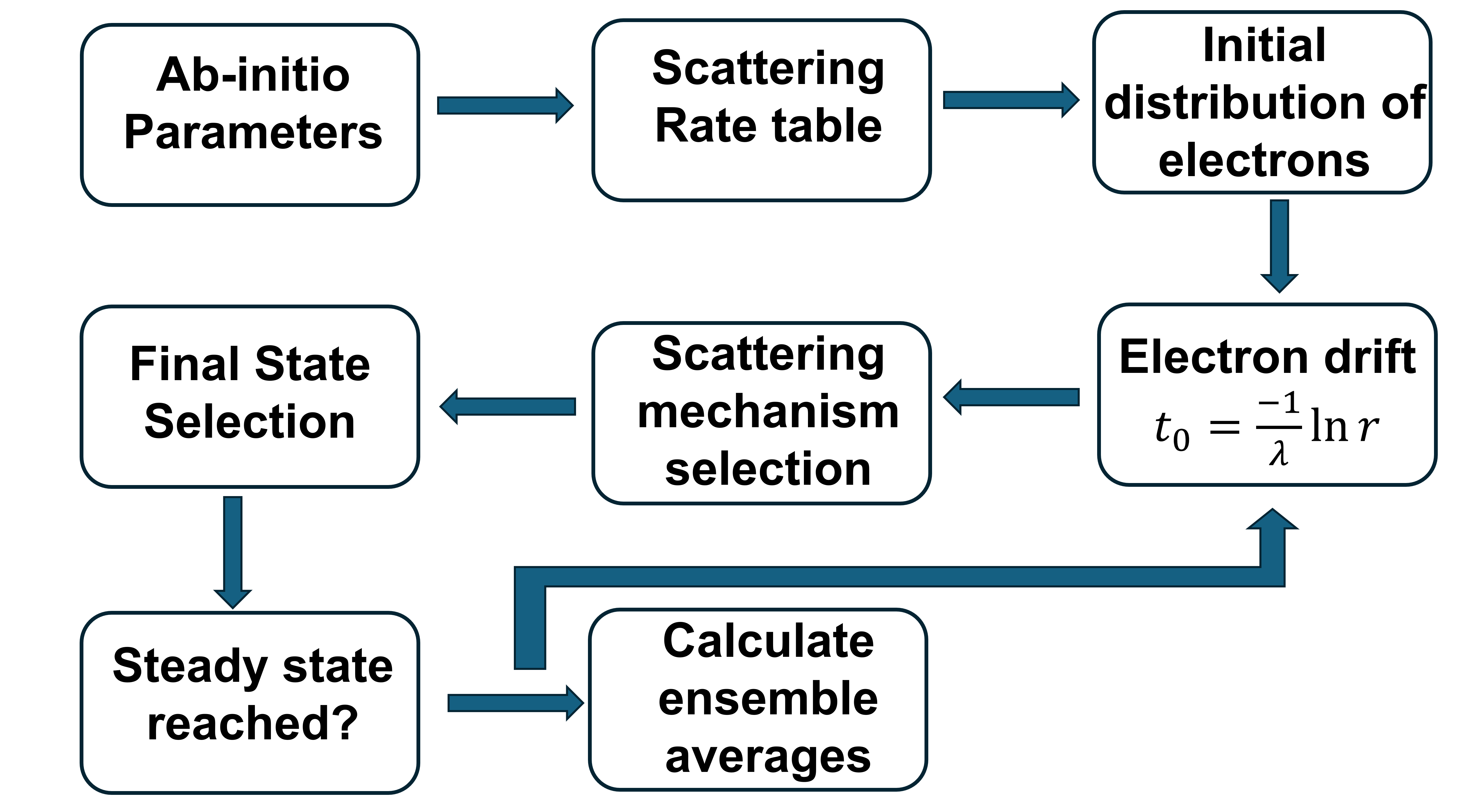}
    \caption{Workflow for the Monte Carlo simulation}
    \label{monteworkflow}
    \end{center}
\end{figure}

In the low-field electron transport regime, the Rode's method can be used to solve the Boltzmann Transport equation (BTE) to obtain the distribution function and investigate the mobility. However under high-field transport conditions, the electrons can drift far from equilibrium thus violating the underlying principle of the Rode's method that the perturbation is small. Hence, under high-field conditions, the Full Band Monte Carlo (FMBC) method is a widely used method to accurately solve the BTE \cite{lundstrom2002fundamentals,jacoboni1983monte,fischetti1988monte}. The underlying principle is based on simulating the trajectories of electrons through the device under the influence of electric fields and choosing the path of the electrons using random numbers based on the probability of the scattering events. The workflow for the FBMC method is outlined in Fig.\ref{monteworkflow}.

The first step in the Monte Carlo simulation is the calculation of the electronic and phonon band structures as well as the scattering rates. As discussed earlier, in this work we have performed first principles calculations and used \textit{ab-initio} calculated parameters.  Then using these parameters, the main workflow of the Full Band Monte Carlo simulation starts which is explained in the next section. 
\subsection{Initial electron distribution and Free Flight}
The initial energies of the electrons are calculated as a room temperature Maxwellian as -3/2 k$_B$T ln(r) where r is a random number.
Once the initial energy is calculated, the initial wavevector of the electron is calculated using a process similar to that of final state selection process which is discussed in the later section. Under the influence of an electric field, the electron is allowed to drift for a free flight time of t$_o$=$\frac{-1}{\lambda}$ ln(r) \cite{kunikiyo1994monte}, where $\lambda$ is the the maximum possible scattering rate from a given band at any k point. The free drift of the electron is governed by a classical picture where $\vec{k_{new}}$= $\vec{k_{new}}$ - $\frac{e\vec{E}t_o}{\hbar}$\cite{kunikiyo1994monte} , where k is the crystal momentum and $\vec{E}$ is the applied electric field. In the methodology of the FBMC method, all the random numbers are denoted by r, but they are generated uniquely  and different from each other. 
\subsection{NSR and Scattering Mechanism Selection}
The next step in the algorithm  after the free flight of the electrons is to select the scattering mechanism the electron encounters. To select the scattering mechanism, a normalized scattering rate table is formed  at each k point and for each band from the \textit{ab-initio} calculated scattering rates with information about the phonon mode, band involved and the nature of the scattering mechanism. Any given scattering rate is normalized in the table in the following way \cite{hess2012monte}:
\begin{equation}
    NSR(\nu)= \frac{\sum_{i=1}^\nu S^{i}_{m}(k)}{max_k \sum_{i=1}^{N^m}S_{m}^i(k)}
\end{equation}
Here m is the given band and $\nu$ is the given mechanism, and N$^m$ is the total number of scattering mechanisms. Next, a given scattering mechanism is selected stochastically using a random number r  such that NSR($\nu$) > r > NSR($\nu$-1). In the other scenario, the electron is considered to be self scattered if NSR(N$^m$) < r and the final state of the electrons remains unchanged. 

\subsection{Final State Selection}
Now once the scattering mechanism is selected with information about the bands involved, phonon mode involved, and the nature of the scattering mechanism, the next step is the calculation of the final state of the electron\cite{kunikiyo1994monte}, which is the most computationally intensive part of the simulation. First, all the k points in the final band n is shortlisted which satisfies the energy conservation $\delta(E_{mk} \pm max_q(\omega_q^i) -E_{nk_j}) $. Here + is for absorption, - is for emission, $E_{mk}$ is the initial energy of the electron, $E_{nk_j}$ are the possible  final state energies of the electron and $\omega_q^i$ denotes the phonon frequency. In the first shortlisting, we ignore the full phonon dispersion and just use the max$_q$($\omega_q^j)$ for the energy conservation to reduce the computational time. In the next step, the full phonon dispersion is used and strict energy and momentum conservation is applied to the k$_j$ points selected to further narrow the list. After this, now one final k$_j$ point is picked stochastically using a random number based on the product of |g(k$_i$,k$_j$ - k$_i$)$^\nu$|$^2$ , which is the EPI elements, and the local density of states (LDOS) at that point \cite{fischetti1993monte}, where g denotes the EPI elements  This step is done carefully taking into consideration the short range or long range nature of the scattering mechanism, which determines the EPI elements at shortlisted k and q point. Once that final k$_j$ is selected , the electron will basically belong to the small cube represented by k$_j$. Now to find the exact final wavevector, the cube is divided into six tetrahedra and the one of the tetrahedra is selected that has more LDOS contribution to the final electron energy $\epsilon_{mk} \pm \omega^i_{k_j-k_i}$. Now finally we identify the equi-energy surface intersecting the selected tetrahedron because that is where the final state of the electron would lie. The shape of the equi-energy surface could be a triangle or quadrilateral. If the shape is quadrilateral, it is divided into two triangles and again one of them is chosen stochastically using a random number based on the area. Once we have a triangular equi-energy surface, the final state of the electron is given as \cite{kunikiyo1994monte,dolgos2013full}:
\begin{equation}
    k_f=r_1[(1-r_2)a+r_2b]
\end{equation}
where r$_1$, r$_2$ are two random numbers and a, b are any two edges vectors of the triangular surface. 

\section{high-field Transport in AlGaN}
\begin{figure*}[t]
\begin{center}
    \includegraphics[width=1\textwidth]{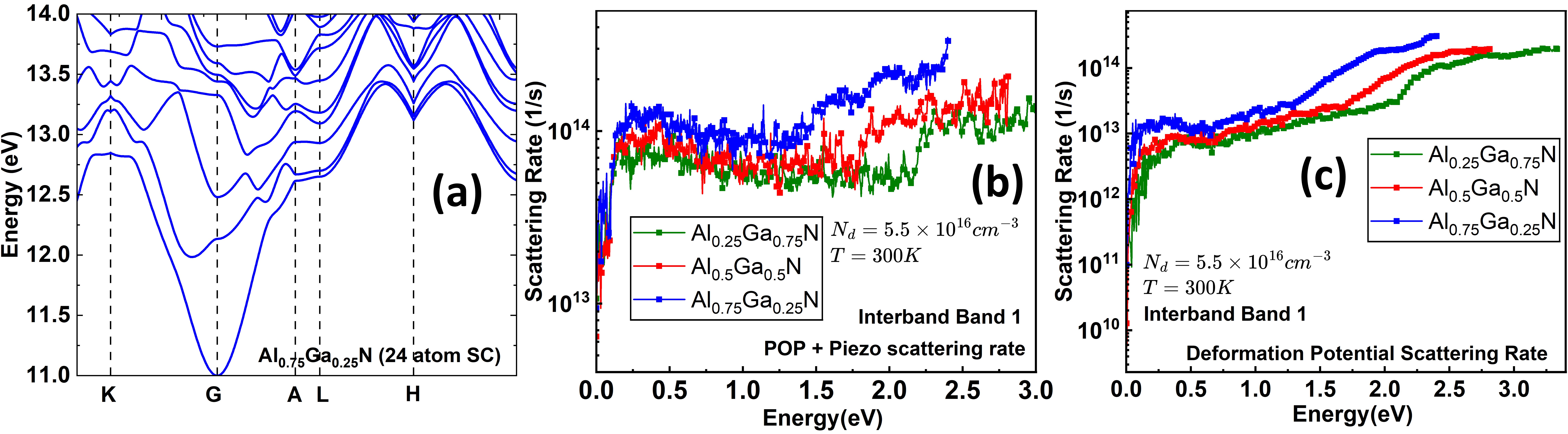}
    \caption{(a) Al$_{0.75}$Ga$_{0.25}$N SC Bandstructure (b) Polar Optical Phonon and piezoelectric scattering rate  and (c)  Deformation potential scattering rate for 25$\%$, 50$\%$ and 75$\%$ Al fraction for intraband and interband Band 1}
    \label{alganscat75}
    \end{center}
\end{figure*}
The FBMC method discussed can be used to study the high-field transport properties like velocity-field profiles and transient dynamics in AlGaN. In this work, we have used a 24-atom SC to study the high-field transport properties in AlGaN. As also discussed in our previous work\cite{datta2024effective}, it is challenging to investigate the transport parameters from the unfolded electronic band structure and phonon dispersion. Thus instead of unfolding the SC properties, in this work, the full supercell is used with multiple bands and multiple phonon modes. Fig.\ref{alganscat75} shows the electronic band structure of the 24-atom SC from where we have included 8 conduction bands till 3eV from the CBM in the FBMC study. 
\\
In our calculations, we include the \textit{ab-initio} calculated scattering rates for the polar optical phonon, piezoelectric, and deformation potential scattering mechanisms for all the 8 bands including the intraband and interband transitions. The donor concentration is assumed to be 5.5e16 cm$^{-3}$ with partial ionization of dopants for the analytically calculated ionized impurity scattering (IIP). The alloy scattering potential is taken as 1.6eV for the three Al fractions of 25$\%$, 50$\%$ and 75$\%$\cite{bellotti2007alloy,jena2003magnetotransport,simon2006carrier,pant2020high}. The ab-initio scattering rates are calculated using Fermi's Golden Rule as shown in Fig.\ref{alganscat75}(b) and (c) for the polar optical phonon, piezoelectric and the deformation potential scattering rate for Band 1 including interband transitions. The Supplementary information (SI) contains further details about the analytically calculated IIP and alloy scattering rates.

For the FBMC simulation, an ensemble of 7000 electrons was used to study the high-field transport properties for electric fields ranging from 20 to 600 kV/cm in the cartesian z direction. The SC based high-field transport method was first verified on GaN by comparing SC and PC based velocity field curves. Please see the details in the SI document.
\subsection{Velocity-field curves and high-field Transport Mechanism}
\begin{figure}
\begin{center}
    \includegraphics[width=0.5\textwidth]{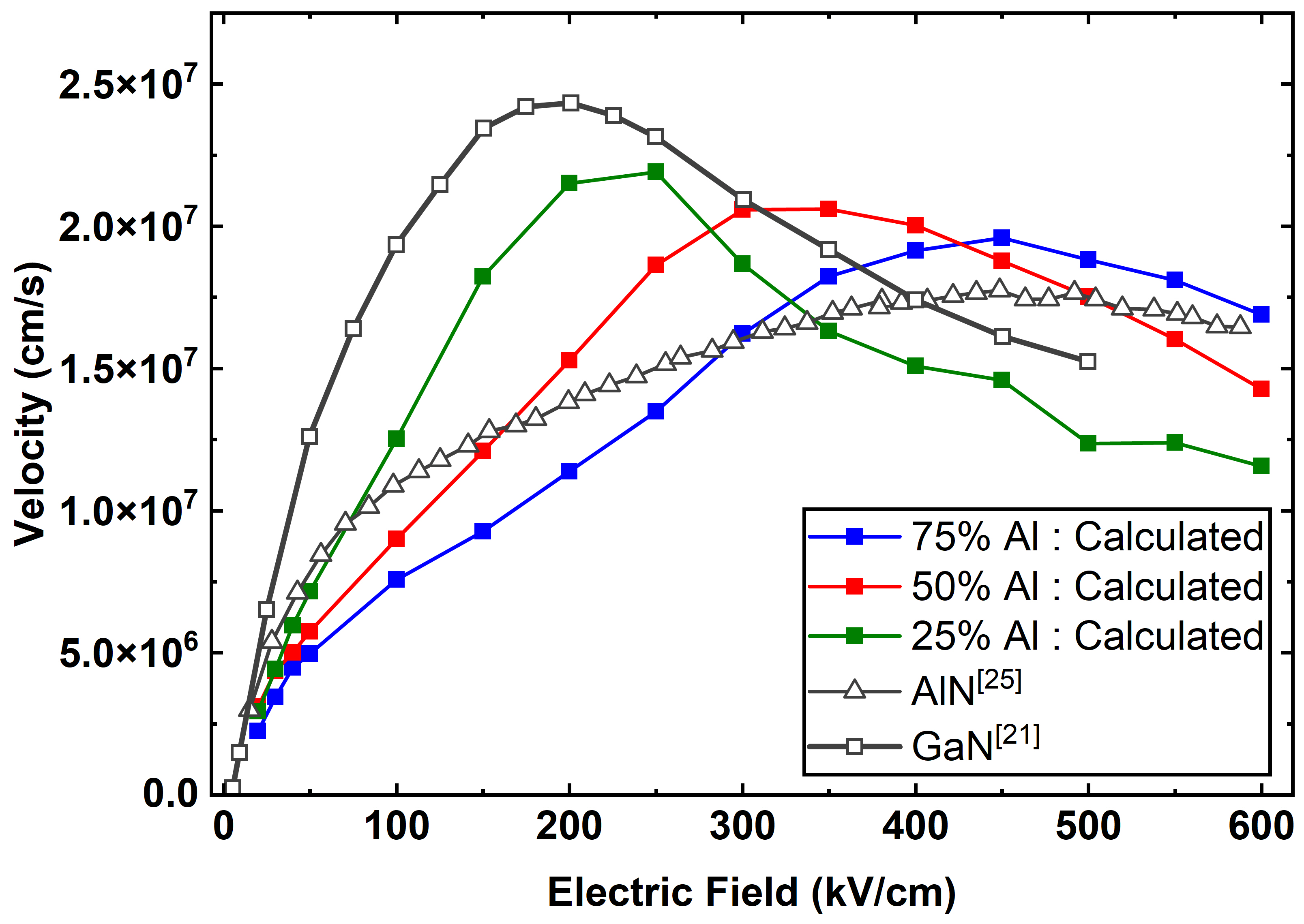}
    \caption{Calculated velocity-field characteristics for Al$_{0.25}$Ga$_{0.75}$N, Al$_{0.50}$Ga$_{0.50}$N, Al$_{0.75}$Ga$_{0.25}$N. The experimental data for the velocity field field characteristics of AlN and GaN are taken from \cite{dyson2015hot},\cite{yamakawa2009rigid}}
    \label{alganvelfield}
    \end{center}
\end{figure}

Using the FBMC method discussed above, the high-field transport in AlGaN alloys for three different compositions is shown in Fig.\ref{alganvelfield}. The electric field in this simulation was applied in the cartesian \textit{z} direction. 
It can be observed from Fig. \ref{alganvelfield}, that a peak drift velocity of 2.1 $\times$ 10 $^7$ cm/s, 2.0 $\times$ 10 $^7$ cm/s and 1.9 $\times$ 10 $^7$ cm/s was observed for the 25$\%$, 50$\%$ and 75$\%$ Al fractions. However even though the peak drift velocity is similar the critical electric field is observed at 250 kV/cm, 350 kV/cm and 450 kV/cm respectively for the 25$\%$, 50$\%$ and 75$\%$ Al fractions. It is noted that the peak velocities are lower than that of GaN and nominally higher than calculated velocity in AlN.

\begin{figure*}[t]
\begin{center}
    \includegraphics[width=0.9\textwidth]{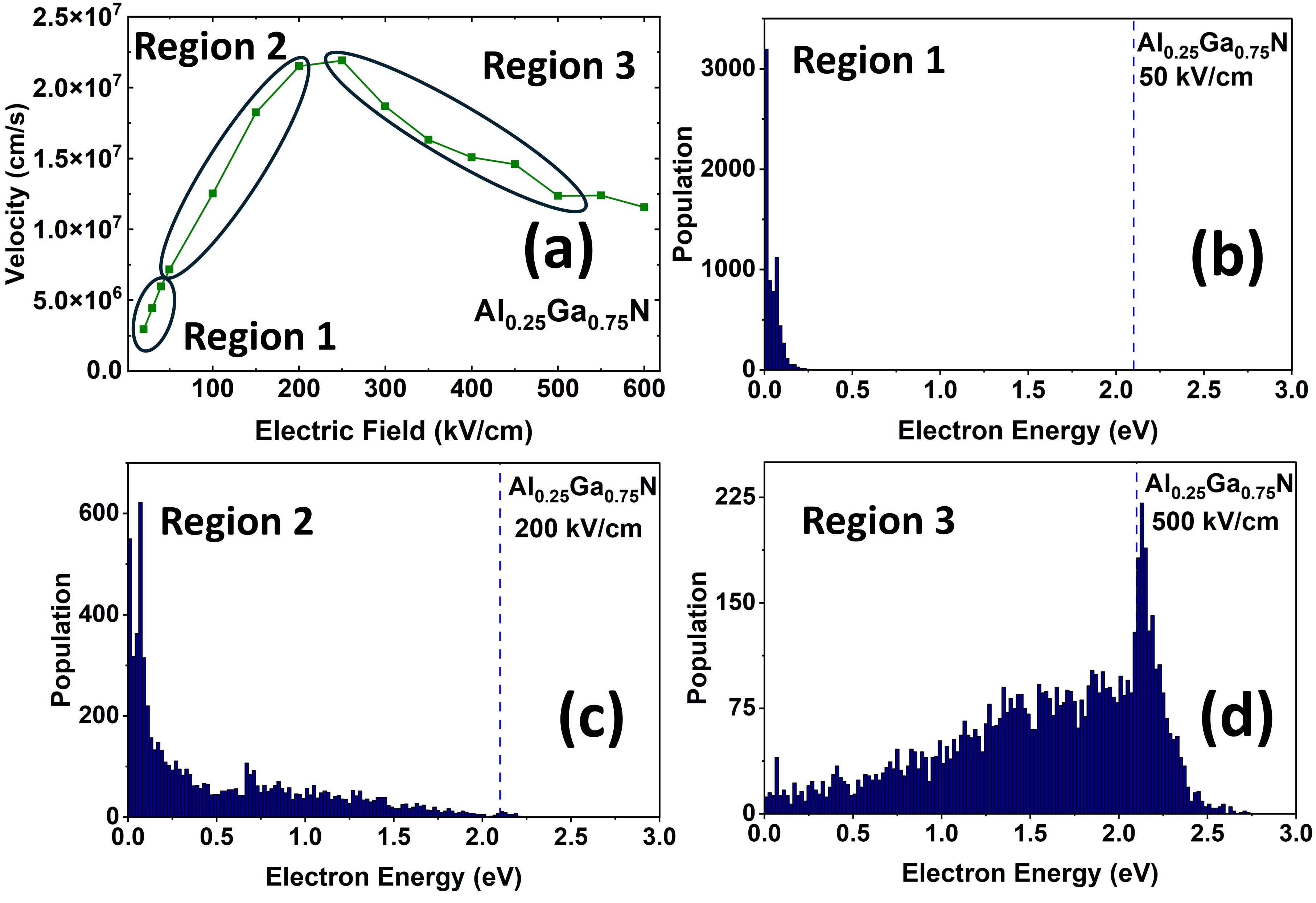}
    \caption{(a) Velocity field profile for Al$_{0.25}$Ga$_{0.75}$N. The energy distribution of the electrons are shown at electric fields of  (b) 50 kV/cm (c) 200 kV/cm and (d) 500 kV/cm. The blue dashed line shows the position of the remote valley energy at 2.1eV for 25$\%$ Al fraction}
    \label{transportmech251}
    \end{center}
\end{figure*}

To further understand these trends, we look at the electron distributions under different conditions.  The transport mechanism is first investigated for Al$_{0.25}$Ga$_{0.75}$N and then compared with Al$_{0.75}$Ga$_{0.25}$N to understand the effects of increasing Al fraction. Fig. \ref{transportmech251}(a) shows the velocity-field profiles for Al$_{0.25}$Ga$_{0.75}$N divided into three distinct regions of transport.
 In region 1, between low to moderate electric fields, the electrons initially gain energy from the electric fields. However, the POP scattering is the dominant scattering mechanisms in this region through which electrons lose energy and the velocity of the electrons increase linearly in this region. The energy of the electrons are also low in this region as shown in Fig.\ref{transportmech251}(b) at 50 kV/cm. As the electric field increases to the moderate to high range in Region 2, the average energy of the electrons starts to increase, with a few electrons transferred to the upper valley as shown in Fig.\ref{transportmech251}(c) at 200 kV/cm. The blue dashed line shows the position of the remote valley energy at 2.1eV from the CBM for the 25$\%$ Al fraction as calculated from the effective band structure (EBS) discussed in our previous work \cite{datta2024effective}. However, in this energy range, the other scattering mechanisms like the deformation potential scattering also starts to play an important role in removing the energy gained by the electrons from the electric field. Thus, even though the electron velocity increases, the rate of increase of the electron velocity is lower as compared to Region 1. 
 
\begin{figure*}[t]
\begin{center}
    \includegraphics[width=1\textwidth]{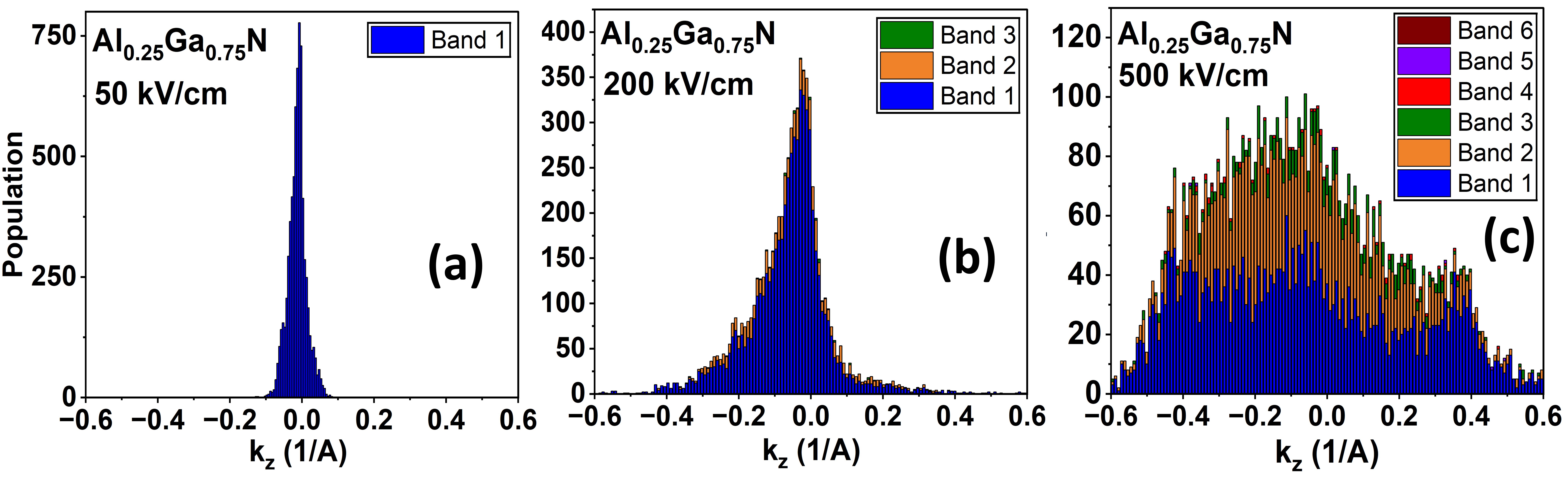}
    \caption{k$_z$ distribution of the electrons in the steady state at electric fields of (a) 50 kV/cm (b) 200 kV/cm and (c) 500 kV/cm for Al$_{0.25}$Ga$_{0.75}$N. The colored bars shows the population of the electrons in various bands.  }
    \label{transportmech25kz}
    \end{center}
\end{figure*}

 Beyond a certain critical electric field, the velocity starts to decreases which is our 3rd region of interest. In this region, the electrons have gained enough energy to transition to the upper valleys as shown in Fig.\ref{transportmech251}(d) at 500 kV/cm. Since the effective mass of the upper bands are higher, the large population of the electrons in the upper valleys essentially decrease the overall drift velocity of the electrons and a Negative Differential Resistance (NDR) like behavior is observed. Fig.\ref{transportmech25kz} shows the distribution in k space as well as the band occupancy of the electrons once the steady state is achieved for electric fields of 50 kV/cm, 200 kV/cm and 500 kV/cm applied in the +z cartesian direction. The near symmetric distribution at 50 kV/cm gradually shifts towards the -z direction with increasing electric fields. It is also observed that with increasing electric fields, the electron population increases gradually in the higher bands. This is consistent with the fact that, with increasing electric fields, the electrons can gain more energy to transitions to higher energy bands. The net velocity of the system is due to the overall contribution of all the electrons distributed in the k space which is evaluated as $v_z=\frac{1}{N} \sum_{i=1}^{N} v_{i,z}$. 

\begin{figure*}[t]
\begin{center}
    \includegraphics[width=1\textwidth]{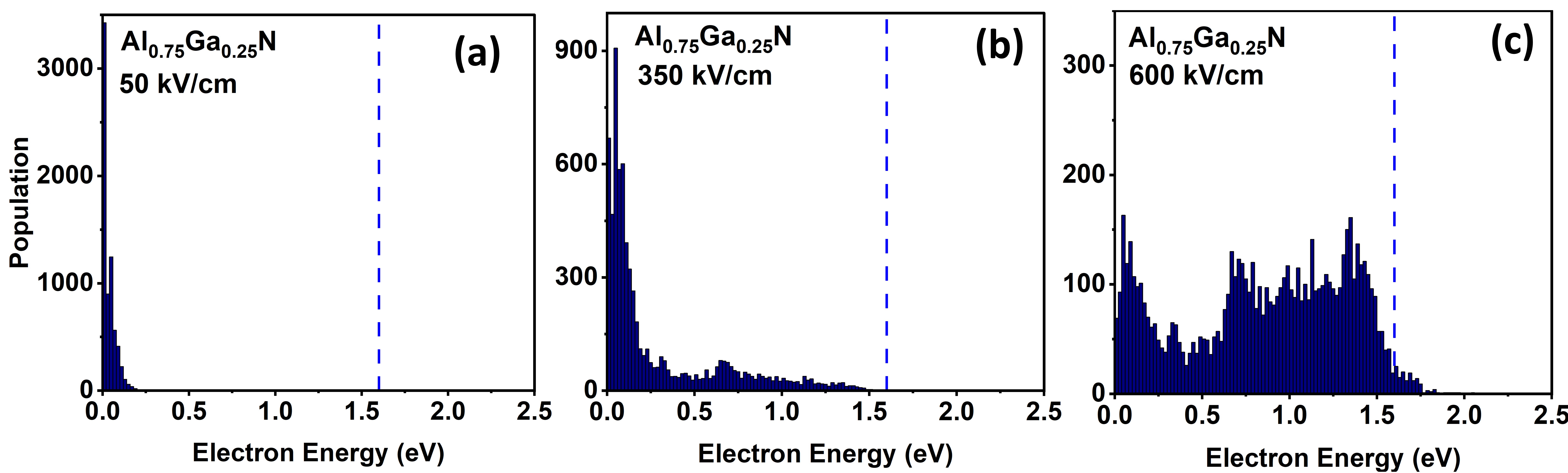}
    \caption{The energy distribution of the electrons is shown at electric fields of  (a) 50 kV/cm (b) 350 kV/cm and (c) 600 kV/cm Al$_{0.75}$Ga$_{0.25}$N. The blue dashed line shows the position of the remote valley energy at 1.6eV for 75$\%$ Al fraction}
    \label{transportmech75}
    \end{center}
\end{figure*}

\begin{figure}
\begin{center}
    \includegraphics[width=0.5\textwidth]{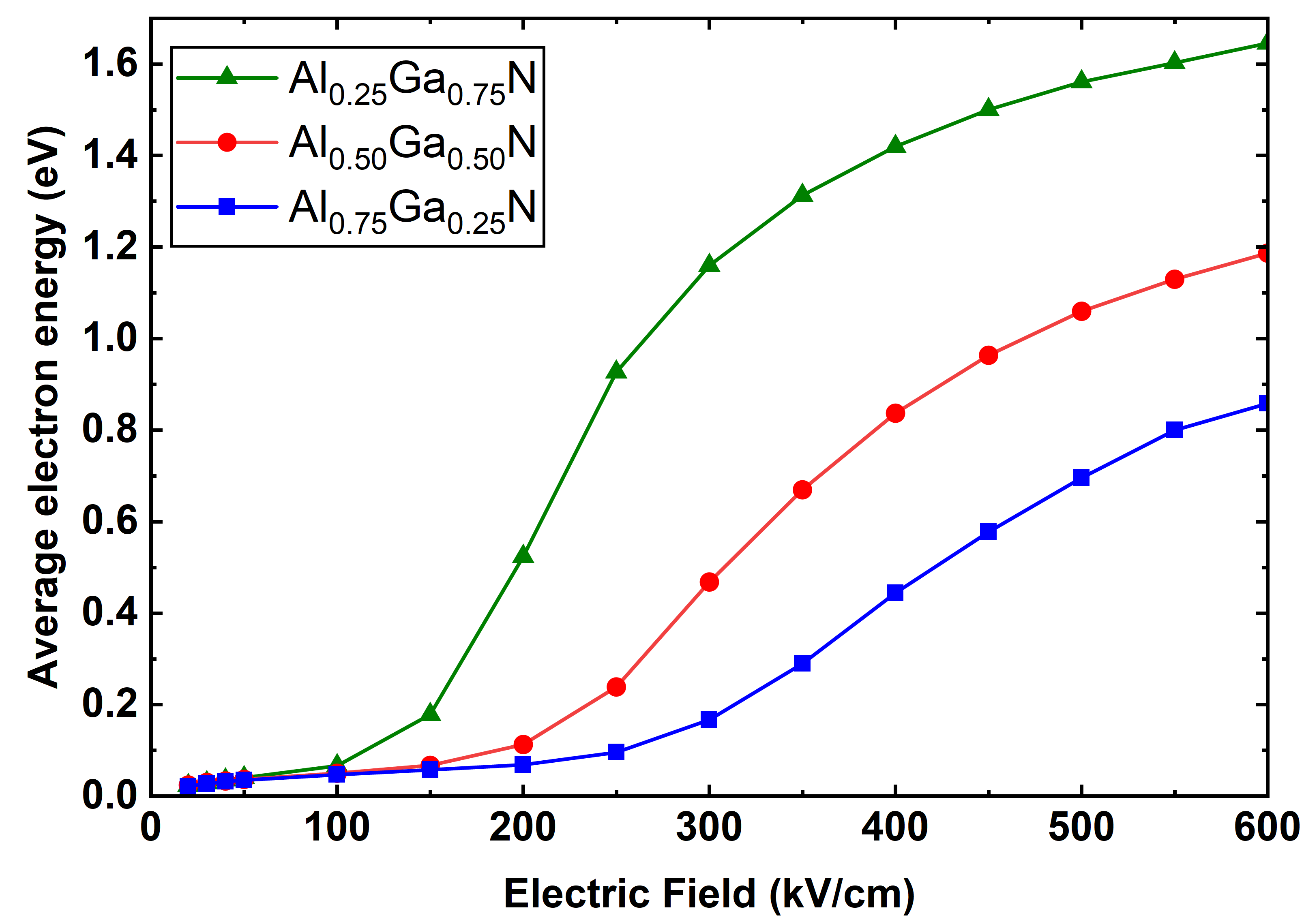}
    \caption{Average electron energy versus electric field for all three Al fractions}
    \label{averageelectronenergy}
    \end{center}
\end{figure}

Next the high-field transport mechanism of Al$_{0.75}$Ga$_{0.25}$N is compared to that of Al$_{0.25}$Ga$_{0.75}$N. Fig.\ref{transportmech75} shows the energy distribution of the electrons at three electric fields of 50, 350 and 600 kV/cm for Al$_{0.75}$Ga$_{0.25}$N. The transport mechanism is similar to the one observed for the 25$\%$ Al fraction. However, the difference in the critical electric field of the system is related to the overall scattering rate . From Fig.\ref{alganscat75}(b) and (c) it can be observed that the POP, piezo and DP scattering rates increases with increasing Al fraction. In case of alloy scattering, even though the scattering potential is same, the scattering rate increase with increasing Al fraction, due to higher effective mass of the electron. Thus due to the high overall scattering rate of the system, the electrons now need higher electric fields to gain enough energy and transition into the upper valleys. This can be clearly observed from Fig.\ref{transportmech75}(c), that even at 600 kV/cm , the electron population in the upper valleys is lower as compared to the 25$\%$ Al fraction case. The lowest remote valley energy for the 75$\%$ Al fraction is 1.6eV from the CBM as calculated from the EBS discussed in our previous work \cite{datta2024effective}. This results in a higher critical electric field for the 75$\%$ Al fraction as compared to the 25$\%$ Al fraction. The SI contains further details about the velocity-field profiles, energy and k$_z$ distribution for the  75$\%$ Al fraction as well as 50$\%$.  
\\
To summarize and visualize the overall picture of the high-field transport mechanism, the variation of average electron energy with the electric field is compared for all the three Al fractions as shown in Fig.\ref{averageelectronenergy}. It can be observed that with increasing Al fraction, the average electron energy is lower at the same electric field. This is consistent with the fact that the increase in scattering rates of the system prevents the electron from gaining high energies. The energy distribution plots shown in Fig.\ref{transportmech251} and Fig.\ref{transportmech75} validates this discussion where it can be clearly observed that at similar electric fields, the energy distribution is in the lower energy range for Al$_{0.75}$Ga$_{0.25}$N compared to Al$_{0.25}$Ga$_{0.75}$N. 
\subsection{Transient Dynamics}
\begin{figure}
\centering
\includegraphics[width=0.5\textwidth]{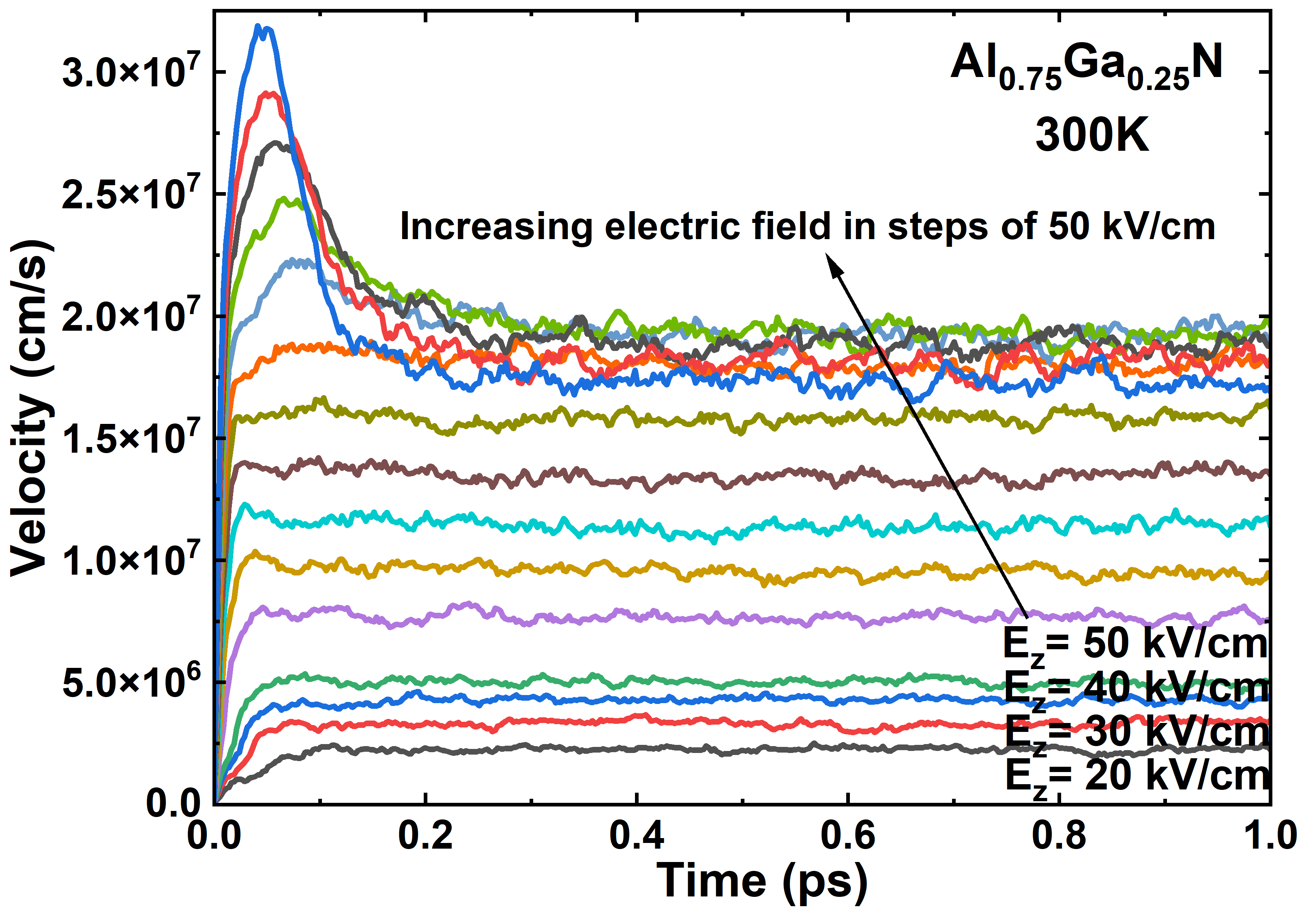}
\caption{The transient electron dynamics of Al$_{0.75}$Ga$_{0.25}$N for a range of electric fields}
\label{transientvel}
\end{figure}

Using the FBMC simulations, it is possible to investigate the transient dynamics of the high-field electron transport. The transient dynamics is a critical aspect of the modern RF electronics where the scaled channel length allows the electron to reach the drain side even without reaching the steady state. Fig.\ref{transientvel} shows the transient electron drift velocity for the 75$\%$ Al fraction. The SI contains the transient electron velocity for the 25$\%$ and 50$\%$ Al fractions. It is observed from Fig.\ref{transientvel}, that for the lower electric fields, the velocity initially rises and reaches the steady state in a short time without any velocity overshoot. However, beyond 450 kV/cm , a significant velocity overshoot is observed. In case of these high-fields, the electrons gain energy and the velocity initially rises sharply. However, at these high-fields, the electron transfers to the upper valleys very quickly and as a result the velocity decreases quickly and reaches a steady state. In order to utilize the velocity overshoot feature and maximize the performance of the RF devices, device designers need to scale the devices to short gate lengths. The overshoot mechanism is similar for all the three Al fractions which is discussed in the SI.

\begin{figure}
\begin{center}
    \includegraphics[width=0.5\textwidth]{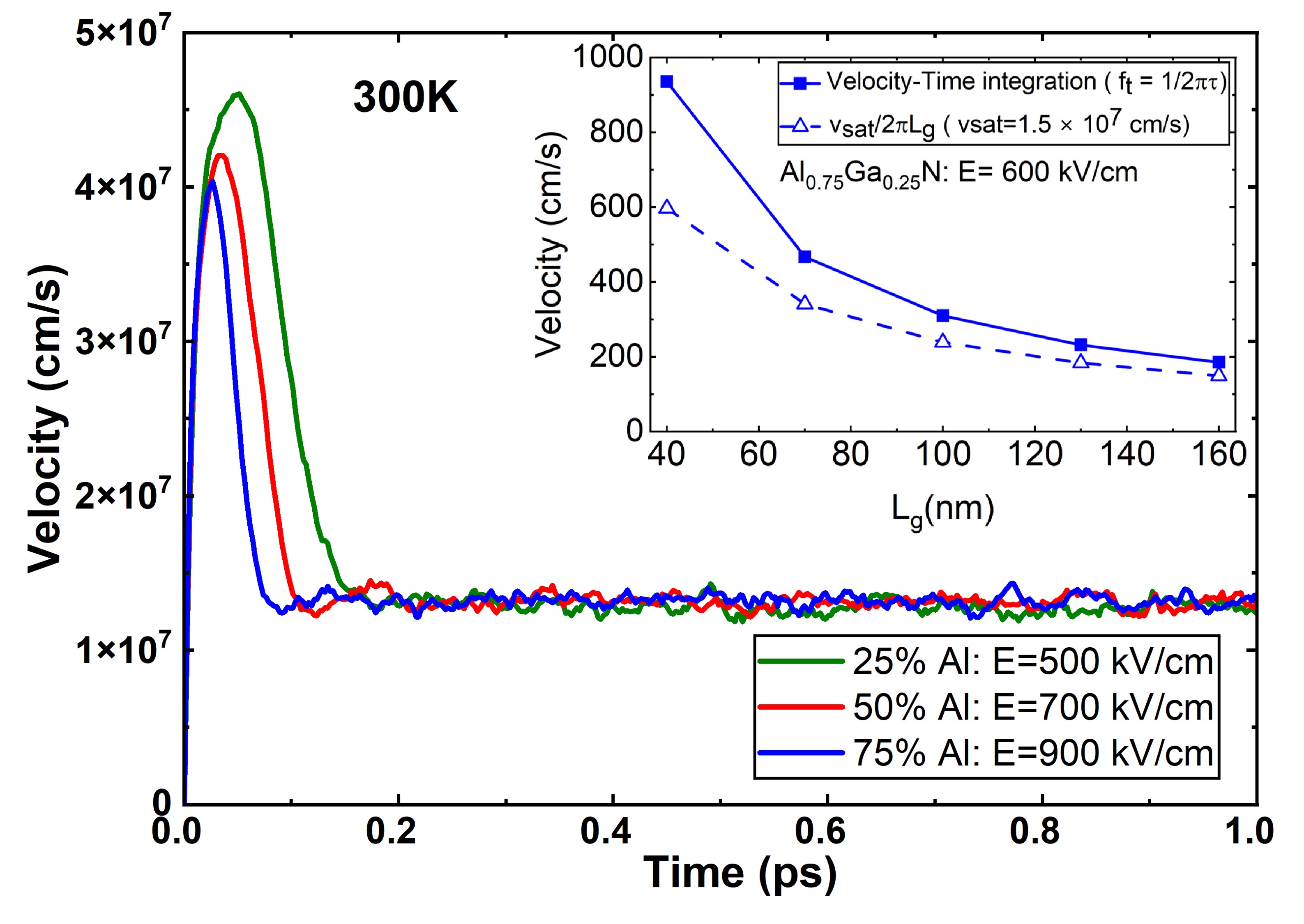}
    \caption{Comparison of the transient dynamics of Al$_{0.25}$Ga$_{0.75}$N, Al$_{0.50}$Ga$_{0.50}$N, Al$_{0.75}$Ga$_{0.25}$N at electric fields of twice the critical electric field for each material. The inset shows the upper limit of f$_t$ calculated from the velocity-transient plots using the velocity-time integration as well as v$_{sat}$/2$\pi$Lg}
    \label{transientcompare}
    \end{center}
\end{figure}

It is also interesting to compare the transient dynamics of the three Al fractions which is dependent on the inter-valley separation energy of the conduction bands. Fig.\ref{transientcompare} shows the transient dynamics of the 25$\%$, 50 $\%$ and 75$\%$ Al fraction at twice the critical electric field for each system to capture the velocity overshoot for all the fractions. The overshoot duration is maximum for the 25$\%$ Al fraction case and decreases gradually as the Al fraction increases. This behavior can be attributed to the fact that the inter-valley separation energy decreases with increasing Al fraction. From the Effective Band Structure (EBS) shown in the SI, the inter-valley separation energy is calculated and tabulated in Table.\ref{intervalley}. 
At very high electric fields, the electrons transfer quickly to the upper valleys because of the lower inter-valley separation energy for the higher Al fractions. The higher effective mass of the upper valleys then slows down the electrons and thus the overshoot duration decreases as the Al fraction is increased. Also,  because of the higher effective masses for higher Al fraction, the peak velocity achieved by the system is also lower for higher Al fractions. 

\begin{table}
\centering
\begin{tabular}{|c|c|}
\hline
\textbf{Material}      & \textbf{Inter-valley seperation (eV)} \\ \hline
\textbf{Al$_{0.25}$Ga$_{0.75}$N} & \textbf{2.1}                          \\ \hline
\textbf{Al$_{0.50}$Ga$_{0.50}$N} & \textbf{1.8}                          \\ \hline
\textbf{Al$_{0.75}$Ga$_{0.25}$N} & \textbf{1.6}                          \\ \hline
\end{tabular}
\caption{Inter-valley separation energy for the three Al fraction calculated from the EBS}
\label{intervalley}
\end{table}
The inset of Fig.\ref{transientcompare} shows the calculated upper limit of f$_t$, which is an important parameter to evaluate the performance of the RF device. f$_t$ is calculated by integrating the velocity-transient plots incorporating the velocity overshoot effects for Al$_{0.75}$Ga$_{0.25}$N at an electric field of 600kV/cm for a range of gate lengths. This result is also compared with the standard formulation of f$_t$ given by v$_{sat}$/2$\pi$L$_g$. The parasitic effects due to the resistance and capacitance are not taken into account in this calculation. The calculated f$_T$ thus provides an upper limit, which highlights the immense potential of these AlGaN based devices. This ideal-limit estimate provides device designers scope to improve device performance and exceed the state of the art technology. 
\subsection{Effect of temperature on high-field transport}
\begin{figure}
\begin{center}
    \includegraphics[width=0.5\textwidth]{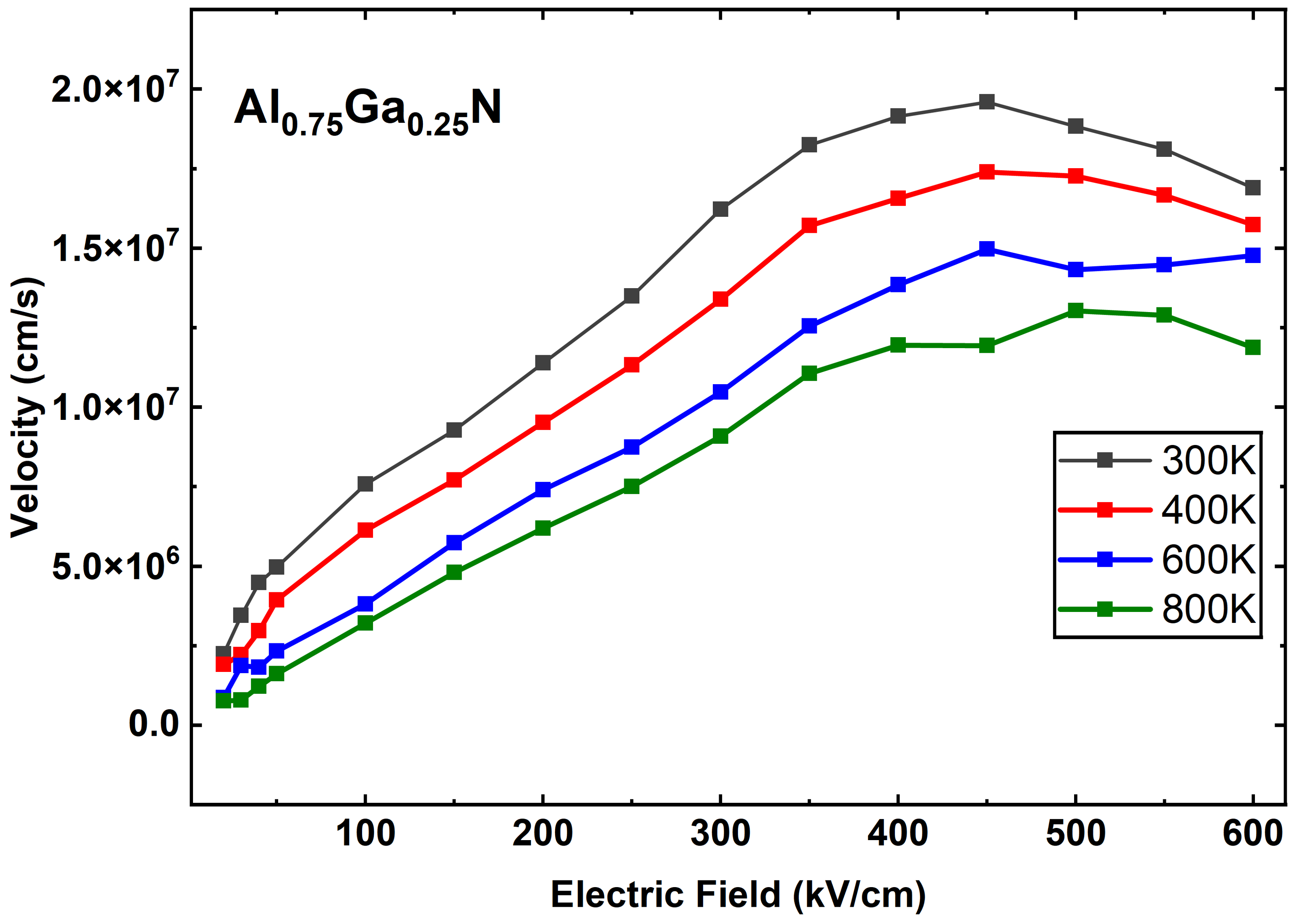}
    \caption{Velocity-field profiles for Al$_{0.75}$Ga$_{0.25}$N for a range of temperatures from 300 to 800K}
    \label{75temp}
    \end{center}
\end{figure}
\begin{figure*}[t]
\begin{center}
    \includegraphics[width=1\textwidth]{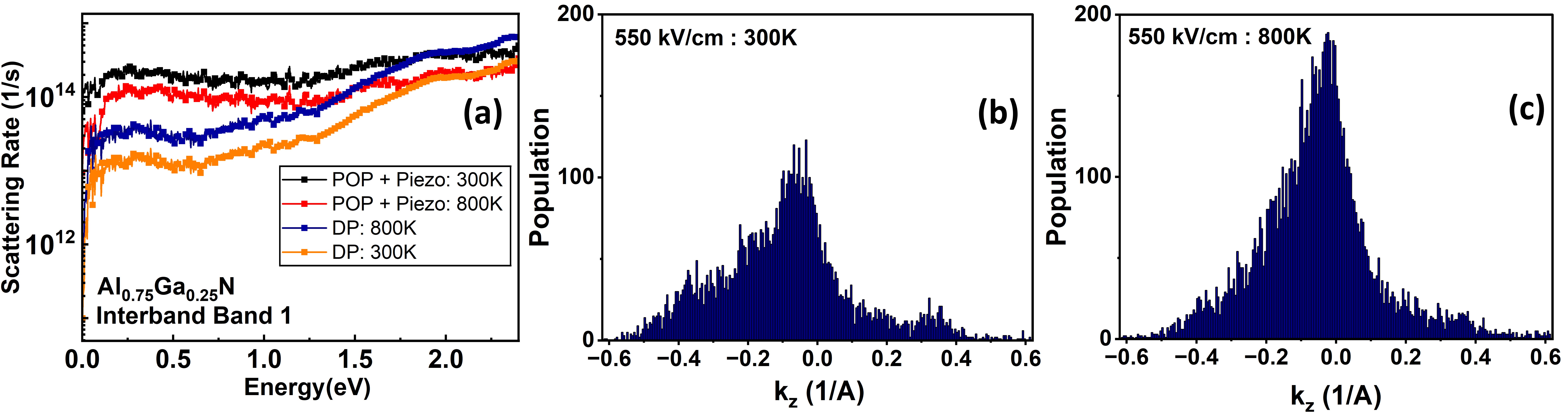}
    \caption{(a) POP+ Piezo (Polar Optical Phonon and Piezoelectric) and the DP (deformation potential) scattering rate at 300K and 800K for Band 1 including intraband and interband transitions. (b) and (c) show the k$_z$ distribution of the electron at 550 kV/cm for 300K and 800K. }
    \label{75temp_kz}
    \end{center}
\end{figure*}
In recent years, there has been increasing interest in the application of UWBG AlGaN for high temperature applications\cite{islam2019heavy,boas2021ionizing}. Thus, it becomes important to study the temperature dependence of the UWBG Al$_{0.75}$Ga$_{0.25}$N for temperatures ranging from 300K to 800K. Fig.\ref{75temp} shows the variation of the velocity field profiles for a range of temperatures for Al$_{0.75}$Ga$_{0.25}$N. It can be observed that with increasing temperature, the velocity of the system reduces and the critical electric field increases as the temperature is increased. The mechanism behind this trend is directly related to the fact that the scattering rates of the system increase with increasing temperature because of the increase in the occupancy of the phonons \cite{hosseinigheidari2026experimental}. Thus the critical electric field increases because electrons now require a stronger electric field to transfer into the higher-energy valleys, which is responsible for the NDR-like behavior.

However, to understand the mechanism underlying the reduction in electron velocity with increasing temperature, it is useful to examine the electron distribution along the $k_z$ direction. Figs.~\ref{75temp_kz}(b) and (c) show the $k_z$ distributions at 550 kV/cm for 300 K and 800 K, respectively. At 800 K, the distribution becomes noticeably narrower than at 300 K. This narrowing can be attributed to the increase in scattering rates at elevated temperatures, as shown in Fig.~\ref{75temp_kz}(a) for band 1 including both interband and intraband scattering rates. Both long-range scattering mechanisms, including polar optical phonon (POP) and piezoelectric scattering, and deformation potential scattering increase with temperature. Consequently, electrons undergo more frequent scattering events reducing their free-flight time and limits the extent to which their $k_z$ vectors can change during transport. As a result, the electron distribution becomes more confined in $k_z$ space. Since the drift velocity represents an ensemble average over all electrons, this narrower distribution contributes to the overall reduction in velocity at higher temperatures. These results provide important insight into the high-temperature transport behavior of UWBG AlGaN and its potential for harsh-environment and high-temperature electronic applications.
\section{Conclusion}
In this work, the high-field transport properties of AlGaN are investigated using the full-band Monte Carlo (FBMC) method using \textit{ab-initio} supercell calcualtion of phonon scattering . The velocity–field characteristics provide critical insight into the saturation velocity, peak velocity, critical electric field, and high-field transport mechanisms in Al$x$Ga${1-x}$N for different Al compositions. The transient dynamics of the system are also examined to understand velocity overshoot effects and their potential role in enhancing RF device performance. Finally, the temperature dependence of transport in ultra-wide-bandgap Al$x$Ga${1-x}$N is studied to determine how elevated temperatures affect its high-field transport behavior.
\section{Supplementary Information}
See the supplementary material for details on convergence of velocity with number of electrons, supercell based Monte Carlo method, and transient dynamics of various Al fractions.

\section{Acknowledgments}
We acknowledge the support from Army Research Office (ARO) under award W911NF-22-2-0163 (Program Managers: Dr. Joe Qiu and Dr. Tom Oder) and Air Force Office of Scientific Research (AFOSR) under award FA9550-18-1-0479 (Program Manager: Ali Sayir), and the Center for Computational Research (CCR) at the University at Buffalo.

\section{DATA AVAILABILITY}
The data and the in-house developed programs that support
the findings of this study are available from the corresponding
author upon reasonable request. The \textit{ab-initio} calculations and quadrupole tensor calculations are performed using the open source software, Quantum Espresso and ABINIT.
\bibliography{aipsamp}

\end{document}